# Paired Cluster Model of High-$T_c$ Superconductivity: Explanation for Pseudogap Critical Concentration, Vortex Core Pseudogap and Possible Stripe Phase in High -$T_c$ Superconductors

J. K. Srivastava*

*Tata Institute of Fundamental Research, Mumbai-400005, India*

(11 April 2005; revised (v2) 19 August 2005)

A new model of high-$T_c$ superconductivity (HTSC) has been developed by us [1-3] which we find is able to explain almost all the HTSC properties, like high $T_c$, origin and nature of NSPG (normal state pseudogap), anomalous superconducting state (SS) energy gap (SSEG) properties, absence of NMR spin relaxation rate coherence peak, existence of NSPG below $T_c$, nature of high-$T_c$ magnetic superconductivity [4], external magnetic field (H) dependence of the NSPG formation temperature T*, etc. The coexistence of NSPG and SS BCS energy gap (EG), below $T_c$, has been predicted by the model before its experimental observation, showing that the observed SSEG is a superposition of NSPG and SS BCS EG. In this paper we show that our model, called the paired cluster (PC) model, is able to explain the pseudogap critical concentration, vortex core pseudogap and possible stripe phase in high-$T_c$ superconductors.

## I. INTRODUCTION

A new model of high-$T_c$ (critical temperature) superconductivity (HTSC), called paired cluster (PC) model, has been developed by us [1-3] which presents a new HTSC mechanism and is able to explain in a natural way almost all the properties of high-$T_c$ superconductors like the high $T_c$, normal state pseudogap (NSPG), superconducting state (SS) energy gap and several other cuprate properties like the absence of coherence peak in NMR spin relaxation rate ($1/T_1$) vs. T variation, high-$T_c$ magnetic superconductivity [4], etc. Extended details of the model, consolidating all our works and giving complete picture, are given in [1] and the readers are strongly recommended to read [1] to get a full appreciation of the model. Actually, some predictions of the models, like coexistence of NSPG and SS BCS energy gap below $T_c$, made years ago (in 1997), have now been experimentally found to be true which grows faith in the model. In this paper we show that our new mechanism, the PC model, can explain the pseudogap critical dopant concentration [5], vortex core pseudogap [6] and possible stripe phase [7] in high-$T_c$ superconductors. However before that, we give below a brief outline of the PC model and its consequences in a way suitable for the present paper.

The PC model is an extension of the earlier proposed cluster phase transition (CPT) model of spin glass (SG) systems which uses the concept of clusters' presence in magnetically frustrated SG lattices [1, 2, 8]. According to the PC model since high $T_c$ superconductors (cuprates) are also magnetically frustrated, magnetic clusters are present in their lattices too. However, unlike CPT model clusters, these (cuprate) clusters occur in pairs. The two pair partners are interpenetrating (Fig. 1) and a spin of one cluster (spin 1) forms a singlet pair, due to the resonating valence bond (RVB) interaction, with a corresponding spin of the partner cluster (spin 2). According to the PC model the conducting electrons (CEs) for T ≥ $T_c$, and both the CEs and the drifting Cooper pairs (CPs) for T < $T_c$, interact with the singlet coupled ion pairs in the cluster and cluster boundaries and this interaction is responsible for the high $T_c$ and other properties of cuprates. A consequence of the CE-, CP- singlet coupled ionic spin pair interaction is to enhance the CE energy, $E_{el}$, by $\Delta E_{el}$, CP energy, $E_{CP}$, by $\Delta E_{CP}$ and the lattice Debye temperature, $\theta_D$, by $\Delta \theta_D$ (as one cools the lattice through $T_{CF}$, $T_c$ where $T_{CF}$ is the temperature at which the PC model clusters are formed). All the cuprate properties can be understood on the basis of these enhancements (scatterings), considered either individually or collectively or in some combination, and the clusters' and cluster boundaries' presence in the lattice. We show this below starting from the critical dopant concentration origin. Though the results given here are for typical parameter values, as discussed in [1-3] they have been checked to be general in nature. The cluster boundaries and the clusters (Fig. 1) have different frustration parameters ($\widetilde{J}$, $\widetilde{J}_0$) since cluster boundaries, compared to clusters, are rich in holes ($Cu^{3+}$ (or equivalently $O^{1-}$) [1, 2]; $\widetilde{J}_1 / \widetilde{J}_{01} > 1$ [1, 2]).

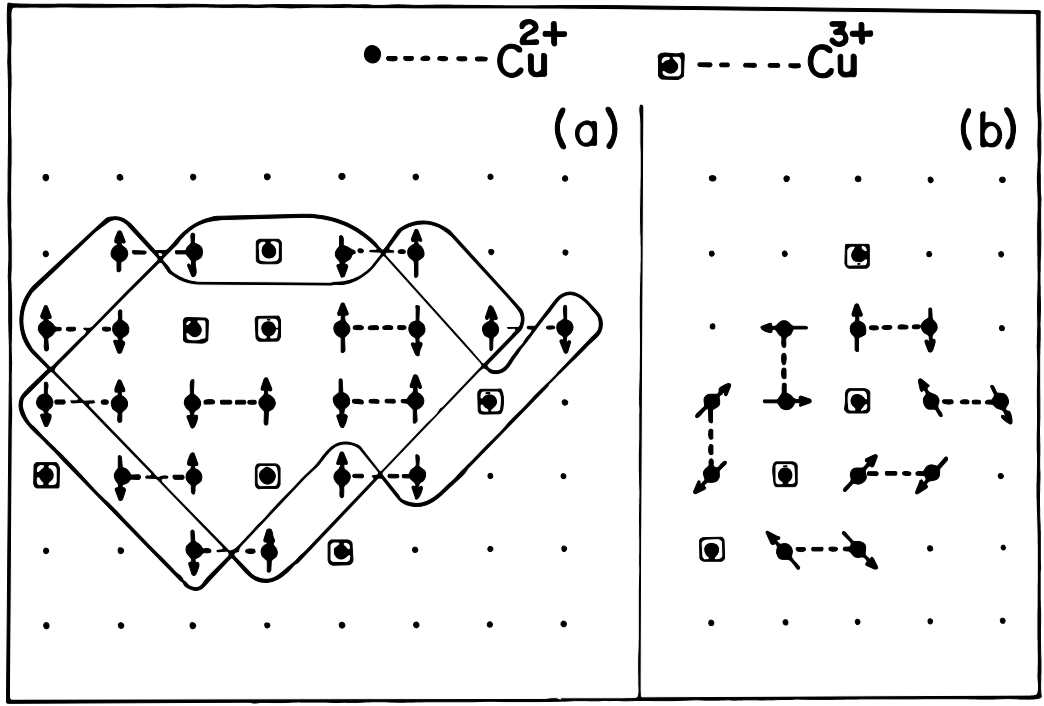

Fig. 1. Schematic representation of (a) magnetic cluster and (b) cluster boundary for a high $T_c$ system. The dashed lines represent the RVB coupling between the spins (arrows) and as explained in the text, clusters exist in pairs due to the combined effect of the frustration and RVB interactions; cluster boundaries, compared to clusters, are rich in holes ($Cu^{3+}$(or equivalently $O^{1-}$) [1, 2]; $\widetilde{J}_1 / \widetilde{J}_{01} > 1$ [1, 2]). In (a), spins of one cluster point in one direction (which is opposite to the partner cluster's spin direction) and in (b), ions (cluster boundary ions) could also be in paramagnetic state (at higher T) [1-3].

## II. CRITICAL DOPANT CONCENTRATION

As discussed in [1-3] the NSPG has been observed in underdoped, optimally doped and overdoped samples, where by optimally doped sample we mean a sample where dopant concentration ($\delta$(as say in $YBa_2Cu_3O_{7-\delta}$) or x (as in $La_{2-x}Sr_xCuO_4$ or $Y_{1-y}Ca_yBa_2Cu_3O_{6+x}$)) is such that the $T_c$ is maximum (i. e. $x(\delta) = x_{opt}(\delta_{opt})$). However in the overdoped region there is a critical dopant concentration, $x_{crit}$ or $\delta_{crit}$,(system dependent) above which NSPG is not observed [5]. We explain below the existence of such a $x_{crit}$ on the basis of the PC model.

As shown in [1-3], in the normal (nonsuperconducting) state ($T_c \leq T \leq T_{CF}$) due to $\Delta E_{el}$ scattering (enhancement) the electronic density of states (DOS), $D(E_{el})$, gets disturbed and there is redistribution of the filled DOS, $D_f(E_{el})$, due to which NSPG (i.e. pseudogap in the redistributed $D_f(E_{el})$, or $D_{ff}(E_{el})$, vs. $E_{el}$ curve ) gets created at the Fermi surface; $N(E_{el})$, the number of CEs at energy $E_{el}$, $= 2 \times D(E_{el})$. The magnitude (size (amplitude, width)), shape of NSPG depends on $\Delta E_{el}$, T (temperature), nature of total DOS [$D_t(E_{el})$], $E_F$ (Fermi energy) and $N_P$, the percentage of CEs for which $\Delta E_{el}$ enhancement occurs [1-3]. In the overdoped region as the dopant concentration increases, more and more $Cu^{2+}$ ions get converted to $Cu^{3+}$ [5] (assuming $Cu^{2+}$, $Cu^{3+}$, $O^{2-}$ picture of the unit cell [1, 2]; the description with $Cu^{2+}$, $O^{1-}$, $O^{2-}$ unit cell picture is equivalent). Due to this $N_P$ gets decreased since $N_P$ is larger if more paired $Cu^{2+}$ ions, with $H_W$ (Weiss field) at their site, are present in the lattice. The $N_P$ decrease is more than just suggested by the $Cu^{2+}/Cu^{3+}$ ratio decrease since as $Cu^{3+}$ increases, $\widetilde{J}_{02}$, defined in [1, 2], gets decreased which enhances the CBIs' (cluster boundary ions') number and diminishes the cluster size i. e. the CIs' (cluster ions') number. Since in the $T_c \leq T \leq T_{CF}$ range CBIs do not contribute to $\Delta E_{el}$ scattering, decrease in CIs' number causes considerable $N_P$ decrease. The $\Delta E_{el}$ is another quantity which gets significantly affected by the $Cu^{3+}$ increase. Our calculations [1, 2] show for the overdoped region a decrease in $\Delta E_{el}$ with increasing x. This happens due to decrease in $H_W$, caused by the $\widetilde{J}_{02}$ decrease, and decrease in $N'_{CE}/N_{CE}$ ratio, defined in [1, 2], occurring since $N'_{CE}$ decreases faster than any $N_{CE}$ decrease owing to the above discussed $N_P$ decrease reason; these findings are consistent with the experimental results [9, 10]. A consequence of $N_P$, $\Delta E_{el}$ decrease with concentration x is that for x larger than a certain concentration $x_{crit}$ (critical concentration), $N_P$, $\Delta E_{el}$ become small enough to make the normal state pseudogap disappear. This is typically shown in Fig. 2 where all descriptions, and parameter values, are same as those of Fig. 7(b) of second reference of [2] (i.e. Fig. 1(b) of fourth reference of [3]), to facilitate comparison, except $\Delta E_{el}(E_F)$ = 100 meV, $N_P$ = 10% (curve a) and $\Delta E_{el}(E_F)$ = 50 meV, $N_P$ = 5% (curve b); (the dotted curve is the total (filled + empty) DOS, $D_t(E_{el})$, [1-3], the dashed curve is $D_f(E_{el})$ and curves a, b are $D_{fr}(E_{el})$ [1-3]; $\Delta E_{el}(E_F)$ = the value of $\Delta E_{el}$ at $E_{el} = E_F$ [1-3]). The pseudogap absence is clearly seen there. This absence gets further confirmed when these curves (a, b) are translated into the tunneling conductance curves [1-3, 9]. Even for somewhat higher values of $N_P$ than the curve a, b $N_P$ values pseudogap is not seen when $\Delta E_{el}$ is small, $\lesssim 5k_BT$; $k_B$ = Boltzmann's constant.

The pseudogap effect absence, i. e. appreciable $\Delta E_{el}$ scattering effect absence, for $x > x_{crit}$ makes these samples behave more like conventional BCS superconductors; this is experimentally seen [5].

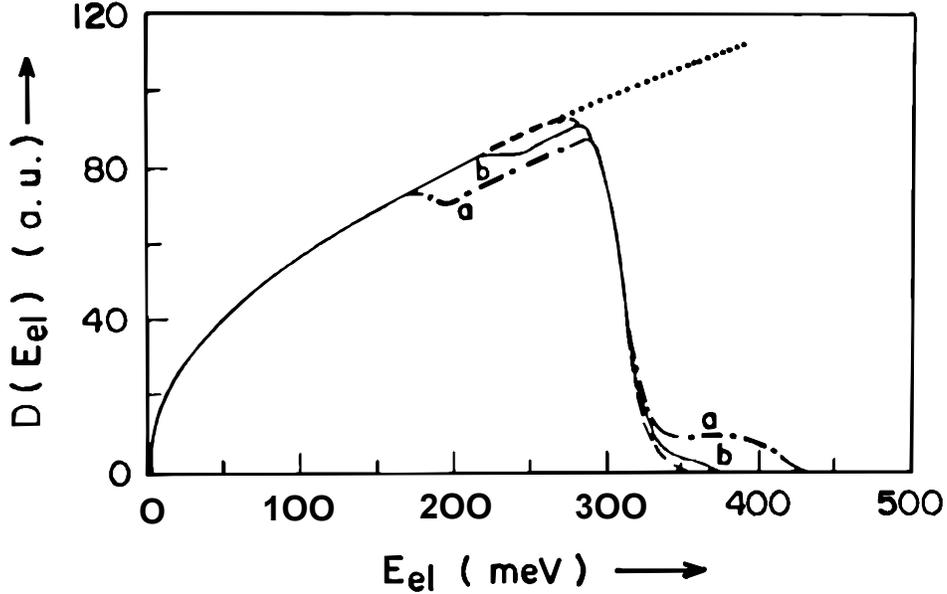

Fig. 2. Dependence of the electronic density of states, $D(E_{el})$, on electrons' energy, $E_{el}$, for $T > T_c$ and x, δ (dopant concentration) > $x_{crit}$, $δ_{crit}$ (critical dopant concentration); $T_c$ = critical temperature, a. u. = arbitrary unit. Details are described in the text.

In the above discussion, as mentioned in the beginning, it has been assumed that the CPs do not exist in the $T_c \leq T \leq T_{CF}$ range. This is because our (PC) model can explain the cuprate experimental results without assuming the CPs' presence above $T_c$. However, there are theories [11] which assume the CPs' existence, without any phase coherence, in the range $T_c \leq T \leq T_{CF}$. We want to point out here that even if such CPs are assumed to be present above $T_c$, our calculations show that the results given in Fig. 2 (and also those given in Fig. 7 of the second reference of [2]) remain almost same. This is since the CPs' number is much smaller than the CEs' number, their (CPs') number varying from ~ 10% of the CEs' number at $T \sim T_c$ to zero at $T \sim T_{CF}$, assuming the CPs' formation at $T_{CF}$, as theories assume [1, 2, 11], and BCS temperature variation for the CPs' number [1, 2]. Thus in $T \sim T_{CF}$ region, CPs' number is very small, ~ 1-2% of CEs' number, and therefore they do not affect the Fig. 2 results any significantly. In the $T \sim T_c$ region, where their number is slightly more, ~ 9-10% of CEs' number, our calculations [1, 2] show that $\Delta E_{CP} \sim \Delta E_{el}(E_F)$ and thus these CPs get scattered in the same way as the CEs. Consequently in the $T \sim T_c$ region too CPs' presence shows no appreciable effect on results. Thus the presence of pseudogap does not prove or disprove the CPs' existence above $T_c$. Similarly in [1-3] we have explained the NMR relaxation rate ($1/T_1$) vs. T anomaly at $T_{CF}$ (i.e. $T^*$ drop; $T_{CF} = T^*$, the pseudogap formation temperature as one cools the lattice from above [1-3]) without assuming CPs' presence above $T_c$. However, this on its own does not rule out such a presence. For instance, if CPs are present above $T_c$, they will only act in cooperation with the effects of our model and further enhance the $T^*$ drop. But we do not favour this explanation since experimental, and also physical, doubts exist on the CPs' presence above $T_c$ [1, 2, 10, 12].

The pseudogap effect absence above $T_c$ does not necessarily mean $\Delta E_{el}$, $\Delta E_{CP}$ scattering effect absence below $T_c$ since, as explained in [1-3], at lower temperatures CBIs can have SG freezing which enhances $N_P$, $\Delta E_{el}$, $\Delta E_{CP}$. Thus three different situations arise depending on whether the CBIs' $T_{SG} \sim T_c$ or $< T_c$ or 0; $T_{SG}$ = SG temperature [1, 2]. For instance, for the $T_{SG} = 0$ case the system behaves like the conventional BCS superconductor right from T = 0K. Similarly for $T_{SG} \sim T_c$, the system will have non-BCS like behaviour only below $T_c$. Some evidence exists in this direction [5]. For more detailed discussion, x vs. $T_{CF}$ and x, T behaviour of experimental data, like say Mössbauer effect, channeling and tunneling conductance data, are needed at least for few cuprates.

## III. VORTEX CORE PSEUDOGAP

The tunneling experiment [6] has shown the presence of pseudogap in vortex core in high-$T_c$ Bi-compound at T << $T_c$ (T=4.2K, H (external field) = 60kOe) which is very similar to the NSPG and changes slowly to the SS gap as one moves away from the vortex core and reaches to a sufficiently separated distance, away from all the other neighboring vortices. The observation of such a vortex core pseudogap has a natural explanation in our (PC) model according to which, as discussed in [1-3], the pseudogap (NSPG) persists below $T_c$ and gets superimposed over the BCS SS gap for T< $T_c$. This happens because $\Delta E_{el}$ scattering is present below $T_c$ too where $\Delta E_{CP}$ enhancement also occurs. Thus according to the PC model, since inside the vortex core system is in the normal (nonsuperconducting) state BCS SS gap is absent and so only NSPG, with parameters (shape, size etc.) modified by low T, H, is observed. Since the magnetic field, which is constant inside the vortex core, decreases, exponentially, as one moves away from the vortex core [13], the pseudogap observed in the vortex core slowly changes from NSPG shape to SS gap shape with increasing distance from the vortex core. This is similar to the NSPG changing slowly to SS gap as T decreases from $T_{CF}$ (=$T^*$) to below $T_c$ where system changes from normal (nonsuperconducting) state to SS. Thus the PC model explanation is consistent with the experimental results [6, 9]. In the normal state ($T_c \leq T \leq T_{CF}$ (=$T^*$)), NSPG changes with T [6, 9]. Similar changes can be observed for the pseudogap inside the vortex core with H.

The observation of NSPG at very low temperatures (T << $T_c$) (in the vortex core) puts doubt on the theories which assume CPs' formation, without phase coherence, above $T_c$ owing to high temperature (T > $T_c$) superconducting fluctuations and associate the NSPG formation to such fluctuations [11]. Since for T< $T_c$ superconducting fluctuations are absent, NSPG at low T (T<< $T_c$) should have not been seen if the above mentioned theories were correct. Also these theories predict the NSPG presence only for underdoped cuprates whereas NSPG has now been observed in optimally doped and overdoped systems also [1, 2, 6, 9]. The recent observation of zero bias conductance peak (ZBCP) in the tunneling conductance spectra of high-$T_c$ Bi-compound and other measurements [14] also argue against the preexisting CPs as the possible cause of vortex core pseudogap [6]. Some connection in the x dependence of NSPG and SS (T < $T_c$) gap is possible since $\Delta E_{el}$ affects both the NSPG and $T_c$, and also since the NSPG superimposes over the BCS SS gap below $T_c$ (i. e. $\Delta E_{el}$ effect is present below $T_c$ too) [1, 2].

## IV. STRIPE PHASE

A large number of theoretical [15-20] and experimental [21-32] works exist concerning the possibility of a stripe phase in cuprates. These theories predict either a macroscopic phase separation [15, 16, 21, 29-31] or a stripe formation [15, 16, 22, 23, 27, 29-31] in cuprates and similar systems. In a macroscopic phase separation the lattice is divided on macroscopic scale into hole full (e.g. $Cu^{3+}$ ($O^{1-}$) [1, 2] full in cuprates ) and holeless (e. g. $Cu^{2+}$ ($O^{2-}$) full in cuprates) regions. Such a phase separation has been observed in insulating $La_2CuO_{4+\delta}$ [15, 16, 21, 29-31]. In a stripe phase we have alternate hole full and holeless regions. For cuprate and similar lattices, which have two dimensional planar structure, theories predict, in the Cu-O or similar planes, $Cu^{2+}$ (or similar spin) full antiferromagnetically ordered regions (domains) separated by one dimensional (or slightly wider ) hole ($Cu^{3+}$ or similar ion) full stripes (domain walls). Such a separation occurs as antiferromagnetically ordered regions expel the holes to minimize region's thermodynamic free energy. Experimentally such a stripe phase has been observed in insulating nickelates, manganites [15, 16, 22, 23, 29-31] and in $La_{2-x}Sr_xCuO_4$, $La_{2-x}Ba_xCuO_4$, $La_{1.6-x}Nd_{0.4}Sr_xCuO_4$ and similar La-Sr-Cu-O based systems for x~1/8 where the system becomes insulating [15, 16, 25, 27, 29-31]. So far such a regularly arranged stripe phase has not been observed in any superconducting cuprate sample. Some stripe phase is indicated by the experiments in superconducting La-Sr-Cu-O based samples (when x is away from 1/8 value) but the stripes there are dynamically fluctuating in space (location) and are not static [15, 16, 24, 25, 28-31]. In other superconducting cuprates indications of stripe phase are not conclusive [7, 15, 16, 26, 29-31]. Physically if all the holes (say $Cu^{3+}$ [1, 2]) stayed in one region and all the spins (say $Cu^{2+}$) in the other region with holes not allowed to jump to the spin region, metallicity would not occur and superconductivity will not be possible. On the other hand if the hole full region is assumed to have some spins in it, i.e. it is not a hole full region but a hole rich region, then conductivity (metallicity) will occur only along the one dimensional hole rich stripe and thus the position (location) of stripes will remain fixed in the space. This is true even if one assumes CPs' formation above $T_c$ in the stripe region and assumes tunneling of CPs, CEs between the stripes to give two-dimensional conductivity. Thus experimentally what we have is, for superconducting cuprates, hole rich and hole poor regions (rather than hole full and holeless regions) with hole rich region stripes either fluctuating in space ( as in La-Sr-Cu-O based systems for x ≠ 1/8) or being not observed at all (as in other cuprates). Experiments do not say anything conclusively about size (width), shape or spatial arrangement of fluctuating stripes. We show below that these results can be understood on the basis of our (PC) model.

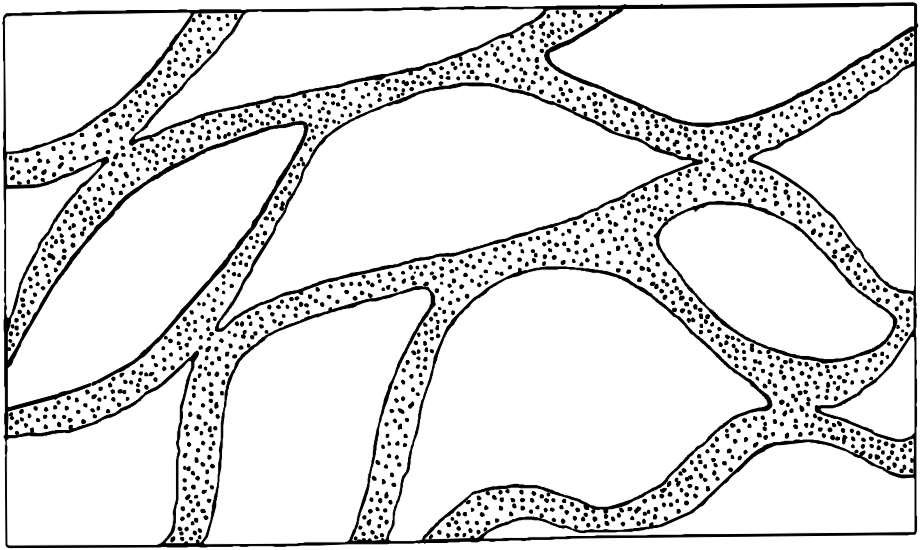

Fig. 3. Schematic representation of possible stripe phase in superconducting cuprates. The cluster boundaries (dotted area) form stripes in Cu-O plane. Details are described in the text.

According to the PC model, we have cluster and cluster boundaries in the superconducting cuprate lattice (Fig. 1). The cluster boundaries are hole ($Cu^{3+}$ ($O^{1-}$)) rich and the clusters are hole poor ($Cu^{2+}$ ($O^{2-}$) rich). This happens because, like the frustrating ions in a SG lattice [1, 8], the doped holes are distributed randomly in the cuprate lattice (at any given instant). Due to this, in the regions where the holes are in excess frustration is large ($\widetilde{J}_1 / \widetilde{J}_{01} > 1$ [1, 2]) resulting in the formation of cluster boundaries, and the spin ($Cu^{2+}$) rich regions (hole poor regions) form clusters ($\tilde{J}_{o2} / \tilde{J}_2 > 1$ [1, 2]). Any preferential segregation of holes, as proposed for insulating lattices [15-20], is not possible for superconducting cuprates since the random $Cu^{2+} \Leftrightarrow Cu^{3+}$ ( or equivalently $O^{1-} \Leftrightarrow O^{2-}$ [1, 2]) charge fluctuation present in their lattice will destroy preferential hole segregation in any region. The cluster boundaries can form stripes in Cu-O plane (Fig. 3) if, as discussed in [1, 2], the frustration is confined within a unit cell along the c-axis and the interplanar coupling is weak, otherwise three dimensional canal (tubular) structure could exist with two dimensional stripe projection on the Cu-O plane. However these stripes (Fig. 3) will have a width, will run randomly in the lattice and will randomly fluctuate in position with time (due to random $Cu^{2+} \Leftrightarrow Cu^{3+}$ ( or $O^{1-} \Leftrightarrow O^{2-}$ ) charge fluctuation). At different instants different types of stripe pattern (arrangement) will be formed in the Cu-O plane and Fig. 3 shows one of the possible stripe patterns which may exist at any particular instant. The details (size, shape, arrangement of stripes) will depend on x, T and sample microstructure [1, 2, 33]. The width of the stripes will vary from place to place, owing to cluster size distribution in the lattice [1, 2], and may even be just one lattice spacing wide at some places [1, 2] or may even not exist at some places (instantaneously broken stripes). For an average 100 Å diameter frustrated area in the Cu-O plane [1, 2], the average stripe width will be ~25 Å if the cluster boundaries and clusters have equal area in the lattice. Thinner stripes will occur if the average cluster boundary area is comparatively smaller (owing to say smaller $Cu^{3+}$ concentration or better lattice microstructure etc.). When a stripe is shared by several clusters, it will be wider. Whereas in most superconductiong cuprates these stripes will be completely randomly located at any instant (complete disorder), in some systems they may have some kind of partial order [17] owing to some specific lattice property like the characteristic buckling of Cu-O plane in La-Sr-Cu-O based superconducting systems [15-17, 25, 29-31]. For partially ordered stripes, some indications may be seen experimentally as fluctuating (dynamic) incommensurate (superlattice) reflections. This seems to be the case in La-Sr-Cu-O based superconducting systems [15-17, 25, 29-31]. Completely disordered stripes will not be detected experimentally. Thus the observed results [15, 16, 21-32] are consistent with the physical picture provided by our (PC) model. Some theories [15-20] too feel the need for bending of stripes, random running of stripes in the lattice, larger width of the stripes, existence of hole rich and hole poor regions ( rather than hole full and hole free regions), instantaneously broken stripes, etc. for understanding the experimental results. The PC model provides a physical basis for the occurrence of the above desired picture. It is also consistent with a recent theoretical work [34] which prohibits the static, charge ordered, stripe formation in superconducting cuprates owing to the presence of next nearest neighbour hopping effect in their lattice. Such a hopping is required for metallicity, and so superconductivity, but it suppresses the static stripe formation.

## V. CONCLUSION

Thus we see that the cuprate properties can be understood on the basis of the PC model. This is true even for those properties which are not specifically discussed here [1-3]. For example, Fig. 4 shows the $D(E_{el})$ vs. $E_{el}$ distribution obtained on the basis of the PC model for $T < T_c$ case [1-3]. The parameters used for calculating various curves (a; b, b′; c, c′) in Fig. 4 are given in [1-3] (e. g. see Fig. 3(b) details of fourth reference of [3]) and a dotted curve is $D_t(E_{el})$, a dashed curve is $D_f(E_{el})$, a full line curve is $D_{fr}(E_{el})$ and a dash-dot curve is the redistributed density of empty states, $D_{er}(E_{el})$. The curve b′ (c′) is obtained by subtracting the unoccupied (empty) side ($E_{el} > E_F + \Delta(T)$, $\Delta(T)$ = SS energy gap width/2) portion of the curve b(c) from the $D_t(E_{el})$ curve. A dip R is seen for curve a (Fig. 4) on P peak side; for the curve b, b′ R appears on P peak side and another dip R′ appears on the Q peak side, and for the curve c, c′ these dips are broad. The location of R, R′ depends on P, Q (gap edges) positions. These results have been experimentally seen [6, 9, 35]. However a variety of explanations have been given for R, R′ existence such as unusually strong pair coupling, d-wave superconductivity, band structure effect, thermal broadening (smearing) effect, mismatch of the pseudogap width and SS gap width at ~ $T_c$, ($\pi$, $\pi$) scattering of electrons in the Brillouin zone, etc. [6, 9, 10, 35, 36]. We see here (Fig. 4) that these dips naturally occur in the PC model in certain circumstances (for certain parameter values) owing to the presence of $\Delta E_{el}$ and $\Delta E_{CP}$ scatterings which, as described before, also explain several other cuprate properties. Thus earlier different explanations have been given for different phenomena, e.g. some earlier explanations for R, R′ are mentioned above, the absence of NMR coherence peak has been associated to enhanced electron-phonon scattering, near $T_c$, owing to the high value of $T_c$ [37], SS stripe phase to the presence of antiferromagnetic correlations [15-20], pseudogap to existence of preformed CPs [1-3, 11] or spin gap presence etc. [1, 2], high value of $T_c$ to nonphononic Cooper pairing etc. [1, 2], and several phenomena remained unexplained like the T dependence of the SS gap width [9], increase in $\theta_D$ at $T^*$ ($\equiv T_{CF}$) on cooling [1, 2], etc. *We see here that all these phenomena can be explained by a common picture, namely $\Delta E_{el}$, $\Delta E_{CP}$, $\Delta \theta_D$, clusters', cluster boundaries' presence, in our (PC) model.* Thus the PC model seems to provide a correct explanation for the HTSC nature and mechanism. The model is flexible enough to incorporate other interactions if needed in some special cases [1, 2].

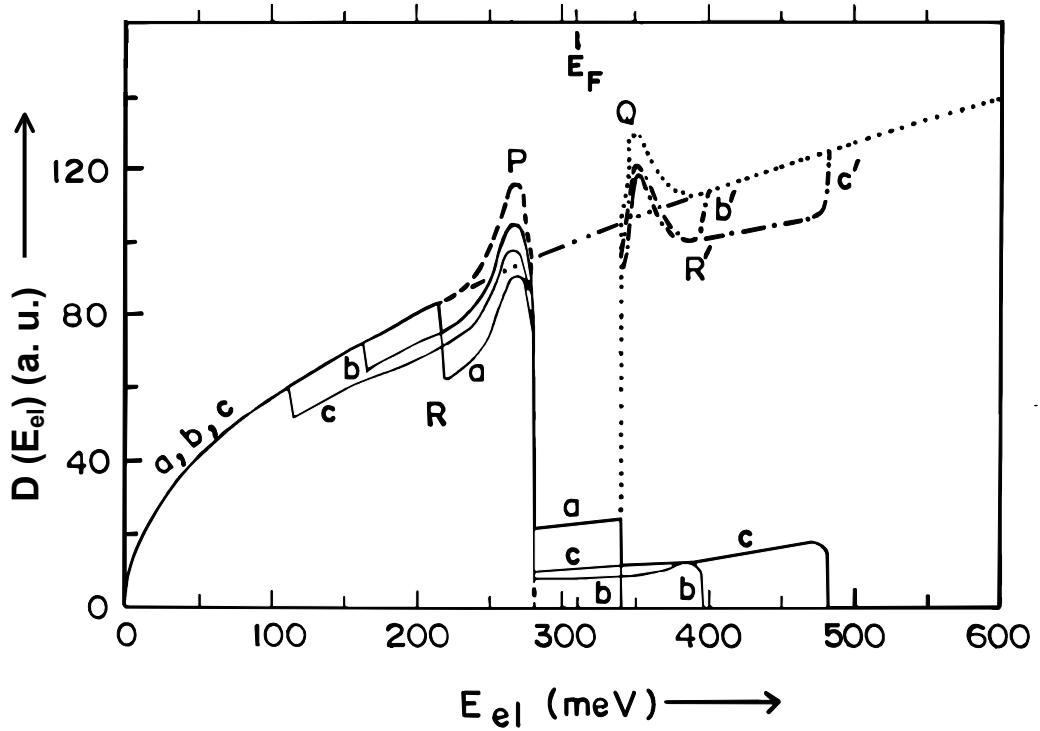

Fig. 4. Dependence of the electronic density of states, $D(E_{el})$, on electrons' energy, $E_{el}$, for $T<T_c$; $T_c$=critical temperature, a. u.=arbitrary unit. Details are described in the text.

It may be mentioned here that the $\Delta\theta_D$ break at $T_{CF}$, referred above and discussed in [1, 2], can be attributed only to the presence of clusters, and cluster boundaries, interacting with CEs, since both the $\Delta\theta_D$ break and the CE- cluster, -cluster boundary, interaction are present in superconducting cuprates only whereas the regular charge ordered stripe phase (resulting from the absence of $Cu^{2+}\leftrightarrow Cu^{3+}$ type charge fluctuation) is present in nonsuperconducting cuprates where no $\Delta\theta_D$ break is observed [1, 2] eventhough the experimental $T_{CF}$ (higher $\Delta\theta_D$ - break temperature) vs. doping concentration, monitored by superconducting samples' study, shows a slow variation (matching with the doping concentration dependence of $T^*$ (pseudogap temperature) seen by several other experimental techniques).Thus $T_{CF}$'s $\Delta\theta_D$ break can not be associated with stripe formation. It may also be mentioned here that the anomalies observed at $T_{CF}$ ($T^*$), $T_c$ are $\Delta\theta_D$ breaks only arising due to a change in lattice ion's r.m.s. vibrational amplitude, and its temperature dependence, and are not a result of any static lattice distortion at $T_{CF}$ or correlated vibration of lattice ions for $T \leq T_c$. This is because these anamolies are observed in Mössbauer f-factor vs. T measurements also as breaks, and change in curve's slope, at $T_{CF}$, $T_c$ (and Mössbauer f-factor senses only the r.m.s. vibrational amplitude of lattice ions and not their static distortion or correlated vibration). In addition, direct measurements of lattice parameters vs. T have also not shown any break at $T_{CF}$, $T_c$ in superconducting cuprates. Only techniques like Mössbauer effect (where Fe substitutes for Cu) or channeling (which can isolate the Cu-O row, or Cu row, selectively) can sense the $T_{CF}$-, $T_c$- $\Delta\theta_D$ break appreciably. Other techniques (bulk or otherwise) may not be able to get appreciable $\Delta\theta_D$- break effect in their studies. Even channeling measurements when carried out for the Y(Er)-Ba row, in $Y(Er)Ba_2Cu_3O_{7-\delta}$, do not find any appreciable $\Delta\theta_D$ break. This also rules out any lattice distortion, like the change in unit cell orthorhombicity, to be the cause of the $\Delta\theta_D$ break since any such distortion, if present, would have affected all the unit cell ions in which case $\Delta\theta_D$ break would have existed for the Y(Er)-Ba row also once it existed for the Cu-O row. It may additionally be mentioned here that whereas the stripe phase theory pedicts a stripe formation temperature very much higher than the pseudogap temperature, with $\Delta'$ (difference between the two temperatures) increasing with decreasing doping concentration, in the underdoped region [11], experimentally wherever stripes (dynamical) have been observed, by experiments like neutron diffraction and NMR, in the underdoped region, the pseudogap temperature and the stripe formation temperature have been found to exist simultaneously; any minute difference in their values, if at all present, may be arising due to the difference in the way the two temperatures are sensed by the experimental technique. This too speaks against the stripe phase theory prediction and so against the $T_{CF}$ - $\Delta\theta_D$ break's association with stripe formation. Similarly it may further be mentioned that the Mössbauer hyperfine field vs. T measurements too favour the earlier discussed low temperature SG freezing of the CBIs.

## VI. SUMMARY

Thus as an overall conclusion we find that the PC model is capable of explaining the HTSC properties either in its present form or in a somewhat modified form. Predictions of the model [1-4] made some years ago, like the coexistence of pseudogap and superconducting state energy gap below $T_c$, are being found true by the experiments now confirming the correctness of the model. *Several other such results too point in the same direction.*

## ACKNOWLEDGEMENTS

Useful discussions with V. R. Marathe (TIFR, Mumbai) and S. M. Rao (National Tsing Hua Univ., Taiwan) are gratefully acknowledged. The author wishes to bring out this work to the attention of the people in this World Year of Physics (2005). Over the period of years, we have not found any experimental observation which is not in favour of the model. Actually, the observed results are found to be a natural consequence of the physical picture contained in the model.

---